**Thermodynamic consequences of molecular crowding in information growth during pre biotic evolution**

Anita Mukherjee* and Arun Kumar Attri**

*Lecturer & Head (Biochemistry), Institute of Public Health & Hygiene, RZ-A-44, Mahipalpur, New Delhi-110038, India

** Professor, School of Environmental Sciences, Jawaharlal Nehru University, New Delhi-110067, India

*Email of Corresponding Author: anitasaumitra70@yahoo.co.in

**Abstract**

The work presented in this paper essentially focuses at providing a scientific theory to explain the growth of information bearing molecules (size and informatio0n contents) without the need of any enzymatic system. It infers the footprints of molecular evolution in the cell interior for a property common to all life forms. It is deducted that molecular crowding is a vital cellular trait common to the all types of cells (primitive or highly evolved). It is argued that this trait is pervasive and must have been incorporated at some stage as a common vital feature of life. If this feature has central importance it must have been part of the pre-biotic information growth of information bearing molecules. The thermodynamic consequences of molecular crowding on the growth of RNA (50-100bp long) in the absence of enzyme system were calculated.

**Introduction**

While talking about evolution, one should start with Charles Darwin who published Origin of species in 1959. The theory gave birth to the new scientific principle and emphasized that the life evolves under selection pressure and has a direction.

Subsequent development not only reinforced the Darwinian evolutionary concept but also shifted the focus of attention on problem of 'Origin of life'. Towards quest of origin of life, there is contribution of enormous scientists e.g. Miller established amino aced syntheses, then synthesis of nucleic acids, and stable polymers such as polypeptides[1], and polyamino acids were reported[2]. Then the problem of self replication and information crisis came i.e. how genetic codes (RNA or DNA) were formed when there were no enzymes (proteins) or how enzymes were formed without RNA the chicken an egg problem. It was though possible to form RNA polymer by activating imidazdlide nucleotide bases in presence of Zn was Pb ions in absence of enzyme, but in this way the polynucleotide sequences of RNA was restricted to 20 to 40 base long[3,4]. For self replication machinery to evolve in the primordial soup, the information contents of RNA in the absence of Enzyme, in terms of the size of the genetic code, should be much larger[5].

**Methodology**

The availability of large information of RNA I crucial as only then the primitive replicate could be translated. These inferences can be drawn by the experiments done[6,7]. In these experiments it was found that the enzyme Qb replicase enzyme in presence of the four energized substrate nucleotide, ATP, GTP, UTP and CTP, and synthesizes RNA, even if no RNA template is present. Q$\beta$ replicase enzyme is in tern coded by 4500 base long genome of Q$\beta$ virus. These results on one hand have shown that evolution of self replication, in the presence of primitive replicate type of enzyme, would have been the sequence preceding evolution of eth cell. But, one the other hand, no explanation came forth, so far, to explain the mechanism of RNA (4500 base long) synthesis, in the absence of the enzyme, for the evolution of the information contents large enough to provide the code for a evolution of the information contents large enough to provide the code for a primitive replicase enzyme system. It means how RNA of size 4500 bas long is made, in absence of enzyme remain a crucial missing piece in the quest to explain prebiotic evolution of enzyme remains a crucial missing piece in the quest to explain prebiotic evolution of self replicating molecular mechanism. In present work we propose to evolve of self replicating molecular mechanism. In present work we propose to evolve a theory which beyond reasonable doubt explains the growth in information, in the absence of enzyme system. If only one assumes that this information crisis somehow was resolved, only then, the rest of the steps leading to the emergence of the self replicating machinery would fall in place. Our theory tries to bridge the gap between 20-40 base long RNA to 4500 base long RNA, in absence of the enzyme the theory is based on thermodynamics of crowded solutions and takes into accounts fundamental traits common to all life form. The proposed theory includes the thermodynamics of molecular crowding and its consequences on the reacting kinetics.

Most of the reaction so far done is in test tube in very dilute condition. The scientific basis and rationale to consider the molecular crowding as central factor comes from the fact that biological reactions are generally performed in very crowded condition, as there are has many macromolecules within cell. The condition inside cell is thermodynamically non-ideal and has significant[8] bearing on chemical kinetics of reaction occurring under such condition[9,10] there are many biochemical reaction which have shown to be increased by molecular crowding e.g. blunt end ligation by DNA ligase increased 1000 times in presence of polyethylene gycol (PEG) crowding[11]. There are many more examples of biochemical processes occurring many fold more in crowded condition like amino acid solubility increases in presence of high concentration of PEG[12] gelling tendency of sickle shaped haemoglobin enhances in presence of non gelling proteins[13], the activity coefficient of cytoplasmic condition of E. coli changed with addition of PEG[14]. Likewise there is a possibility of crowding playing a significant role in increasing RNA size.

In the equation, $\Delta G=\Delta H-T\Delta S$, where $\Delta G$ is change in free energy, $\Delta H$ is change enthalpy, T is absolute temperature and $\Delta S$ is change of entropy,

If there is no change in enthalpy and temperature being same, change in $\Delta G$ is only due to change in entropy. Molecular crowding causes change in entropy thereby increasing the spontaneity of the reaction. We have calculated increases of thermodynamic activity of information bearing molecules (RNA) in presence of polypeptide molecules by using Scale Particle Theory developed by Reiss, Frisch and Lebowitz in 1959, and modified by Gibbons.

**Results**

**A) Chemical Potential**

The calculations of chemical potential was done for varying conditions, shown in table, for different sized of polyp varied from 2,4,10,20,50, and 100 base pairs. The basis of selecting size below 100 base pairs falls within the con-evidence for the growth f RNA, in the absence of enzyme, under simulated prebiotic conditions[15,16]. The size of polypeptide chain was also varied from 2, 4,10,20,50 and 100 base pairs long. Experimental evidence validates the presence of the large amount of polypeptides (PP), of diverse sizes, during prebiotic molecular evolution[17]. The crowding was assumed to be introduced due to the presence of PP in the given volume V. For convenience, species i, where i represent the RNA, was normalized by dividing in with R.T. Here, R (1.98717 Cal. $K^{-1}$ $Mol^{-1}$) is gas constant and T (300 K) represents the temperature of the reaction conditions. The steps provide the estimation of $\mu$/R.T, which becomes a unit less quantity. At the same time, this ratio represents the chemical potential of species I at constant T for a given set of calculations. The changes in $\mu/R.T.$ with corresponding variation in volume occupancy in a given volume were presented in graphic form in figures 24 to 60 (even numbered figures only). The plotted ratios of chemical potential of RNA were normalized. In our calculations, the temperature value of 25 C (298 K) has been used, unless specified otherwise. For the purpose of the calculations the concentration of RNA was kept constant at 100 $\mu g/ml$. This concentration was chosen on the basis that approximately 1% of the cellular weight is confined with nucleic acid[18]. Following salient features from these graphs are inferred from molecular crowding experiments using MathCAD and Hamog software.

1. $\mu/R.T$, which is = $\ln(\gamma_i)$, increased with the increase in fraction of volume occupancy exponentially, for all the cases. Significant change in values occurs on account of the variation in the size of RNA and polypeptide. The starting value of $\phi_i=0$, corresponds to the limiting case representing the ideal condition (absence of molecular crowding effect). The increase in $\mu/R.T$, with increasing $\phi_i$ showed that the reaction conditions are favored to proceed to the right (the size increase of RNA). In other words the polymerization of RNA, with activated ends in the presence of activated nucleotide bases, will be favored with the increase in the molecular crowding in the given volume.

2. It is evident that for the conditions, where the combination of 2 base paired RNA and dipeptide were present in the given volume (Figure 24). At $\phi_i= 0.250$ the value of $\phi_i/R.T.$ is 6.116. This value is comparable with maximum value of $\phi_i/R.T$ for the case where the initial degree of polymerization for the RNA and polypeptide was same, except for the case where RNA and polypeptide had 50 and 100 base.

3. The value of $\phi_i$/R.T. was lower for case where the initial size of RNA and polypeptides were in following combinations ;( 2:4), (20:50) and (20:100). The observed values at $\phi_i$ =0.254 varies from 4.33 to 5.171.

4. For RNA: Polypeptide size combinations; (4:2), (20:10), (20:2), (20:4),(50:2), (50:4), (50:10), (50:20),(100:4),(100:10)(100:20) and (100:100); at, $\phi_i$ =0.254 the $\mu_i$/R.T. value ranges between 10 to 114.05, behavior was observed at $\phi_i$ =0.036, 0.0730, 0.103 and 0.186.

5. The lower values of $\mu_i$ /R.T. were observed for the RNA: Polypeptide size combinations; (2:100), (2:10), (2; 20), (2:50), (4:10), (4:20), (10:50) and (10:100); at, $\mu_i$/R.T. ranges between 1.35 to 3.9 at $\phi_i$ =0.254 (shown here) and at lower values of $\phi_i$

6. Interestingly, the lowest value of $\mu_i$ /R.T. = 1.357 was for RNA: Polypeptide combination 2:100 (figure 30). On the other hand the highest value for size combination 100:2 ($\mu_i$/R.T=176.51).

7. The value of $\mu_i$ /R.T at $\phi_i$ =0.254 for different size combination of RNA: Polypeptide is given in table 1.

B) **Standard Free Energy Change**

The standard free energy change normalized to the R.T ($\Delta G_i$ /R.T.) due to molecular crowding, by virtue of thermodynamic relation show a reversed trend to that observed for $\mu_i$ /R.T. $\Delta G_i$/R.T. is plotted as function of volume occupancy, $\phi_i$, figure 25. The function is indicative of the fact that the polymerization of RNA to the larger size, from the initial size, will be favored.

1. The plotted function $\Delta G_1 / R.T$ decreased with increase in the fraction of volume occupancy exponentially, for the values occurred on account of the variation in the size of RNA and polypeptide. The starting value of $\varphi_i = 0$, corresponds to the limiting case presenting the ideal condition (absence of molecular crowding). The polymerization of RNA, with activated ends in the presence of activated nucleotide bases, will be favored with the increase in molecular crowding.

2. It is evident that for the conditions, where the combination 2 base paired RNA and dipeptide are present in the given volume.

    TABLE -1: $\mu_i / RT$ values at $\phi_i = 0.254$ for different initial size of RNA and polypeptide in ascending order.

| RNA | Polypeptide Size | $\mu_i/RT$ |
|---|---|---|
| 2 | 100 | 1.357 |
| 2 | 50 | 1.464 |
| 2 | 20 | 1.818 |
| 4 | 100 | 1.823 |
| 4 | 50 | 1.984 |
| 2 | 10 | 2.415 |
| 4 | 20 | 2.535 |
| 4 | 10 | 3.48 |
| 10 | 100 | 3.616 |
| 10 | 50 | 3.985 |
| 20 | 50 | 4.007 |
| 2 | 4 | 4.33 |
| 20 | 100 | 4.668 |
| 10 | 20 | 5.269 |
| 2 | 2 | 6.116 |
| 4 | 4 | 6.602 |
| 20 | 20 | 7.027 |
| 10 | 10 | 7.507 |
| 4 | 2 | 9.603 |
| 20 | 10 | 10.332 |
| 50 | 100 | 10.769 |
| 50 | 50 | 12.016 |
| 10 | 4 | 15.036 |
| 50 | 20 | 16.66 |
| 20 | 4 | 21.85 |
| 100 | 100 | 21.917 |
| 10 | 2 | 22.354 |
| 100 | 50 | 24.549 |
| 50 | 10 | 25.012 |
| 20 | 2 | 33.507 |
| 100 | 20 | 34.324 |
| 100 | 10 | 51.978 |
| 50 | 4 | 54.352 |
| 50 | 2 | 84.054 |
| 100 | 4 | 114.050 |
| 100 | 2 | 176.51 |